\documentclass[twocolumn,twocolappendix]{openjournal}
\usepackage{float}
\usepackage{xcolor,bm}
\usepackage{amsmath}
\newcommand{\edit}[1]{\textcolor{black}{#1}}

\begin{document} 

\title{Dark Matter Particle Flux in a Dynamically Self-consistent Milky Way Model
}

\author{Lucijana Stanic$^1$, Mark Eberlein$^1$, Stanislav Linchakovskyy$^1$, Christopher Magnoli$^1$,
Maryna Mesiura$^1$, Luca Morf$^1$, Prasenjit Saha$^1$, Eugene Vasiliev$^2$}
\affiliation{$^1$ University of Zurich, Switzerland}
\affiliation{$^2$ University of Surrey, UK}

\begin{abstract}
We extend a recently developed dynamically self-consistent model of the Milky Way constrained by observations from the Gaia observatory to include a radially anisotropic component in the dark matter (DM) halo, which represents the debris from the accreted Gaia-Sausage-Enceladus (GSE) galaxy.  
In the new model, which we call a self-consistent Anisotropic Halo Model or scAHM, we derive distribution functions for DM velocity in heliocentric and geocentric reference frames.  We compare them with the velocity distributions in the standard halo model (SHM) and another anisotropic model (SHM$^{++}$).  We compute predicted scattering rates in direct-detection experiments, for different target nuclei and DM particle masses.  Seasonal dependencies of scattering rates are analyzed, revealing small but interesting variations in detection rates for different target nuclei and DM masses. Our findings show that the velocity distribution of the anisotropic GSE component significantly deviates from Gaussian, showing a modest impact on the detection rates. The peculiar kinematic signature of the radially anisotropic component would be most clearly observable by direction-sensitive detectors.
\end{abstract}

\maketitle

\section{Introduction}
The study of dark matter (DM) presents a fascinating paradox in modern astrophysics. 
Despite comprising five times more matter in our universe, DM remains elusive, evading direct detection and defying precise characterization. 
Its existence is inferred primarily through its gravitational effects on visible matter.
Such observational evidence for DM comes from various sources, including galaxy rotation curves \citep[e.g.,][]{bosma_2023} and the cosmic microwave background \citep{planck_parameters_2018}.
These studies demonstrate the influence of DM on large-scale cosmic structures; however, our understanding of its fundamental nature remains limited (see e.g. recent reviews by \citealt{bertone_hooper_2018} and \citealt{cirelli_2024}).

One of the most striking uncertainties about DM is its mass. Current hypotheses span an enormous range, depending on the DM candidate assumed -- from ultra-light particles with masses as low as 10$^{-22}$ eV/c$^2$ \citep{BEDM_Broadhurst_2020} to massive objects of up to 5 solar masses \citep{MACHO_rodriguez_2014}. 
The lower bound arises from observations of dwarf galaxies, as any lighter particles would possess a de Broglie wavelength too large to be confined within these small galaxies. 
Conversely, the upper limit is often associated with primordial black holes, whose potential presence is investigated through simulations of their gravitational interactions with wide binary star systems in galactic halos.

This vast uncertainty in DM properties poses a significant challenge for detection. Nonetheless, extensive efforts are underway to develop Earth-based experiments capable of revealing its nature. These experiments fall into three main categories \citep{queiroz_2016}:
\begin{enumerate}
    \item Direct detection: The underlying principle of direct detection experiments is that DM particles are assumed to interact weakly with regular matter, occasionally colliding with atomic nuclei or electrons in the detector material. This should produce detectable signals such as heat, ionization, or scintillation light. Detector technologies used include cryogenic detectors producing phonons after interactions, noble liquid detectors producing scintillation light and ionization, and semiconductor detectors detecting nuclear and electron recoils.
    \item Indirect Detection: These methods involve looking for byproducts of DM interactions such as annihilation or decay, which could produce gamma rays, neutrinos, and other standard model particles. These signals are then searched for in astrophysical sources like galactic centers and dwarf galaxies. The measurements are made using gamma-ray telescopes, neutrino detectors, and cosmic ray detectors.
    \item Collider Searches: These aim to produce DM particles by smashing protons or heavy ions at high energies. If DM particles are created, a significant imbalance of momentum and energy should be noticed.
\end{enumerate}

\begin{figure*}
    \centering
    \includegraphics[width=1\linewidth]{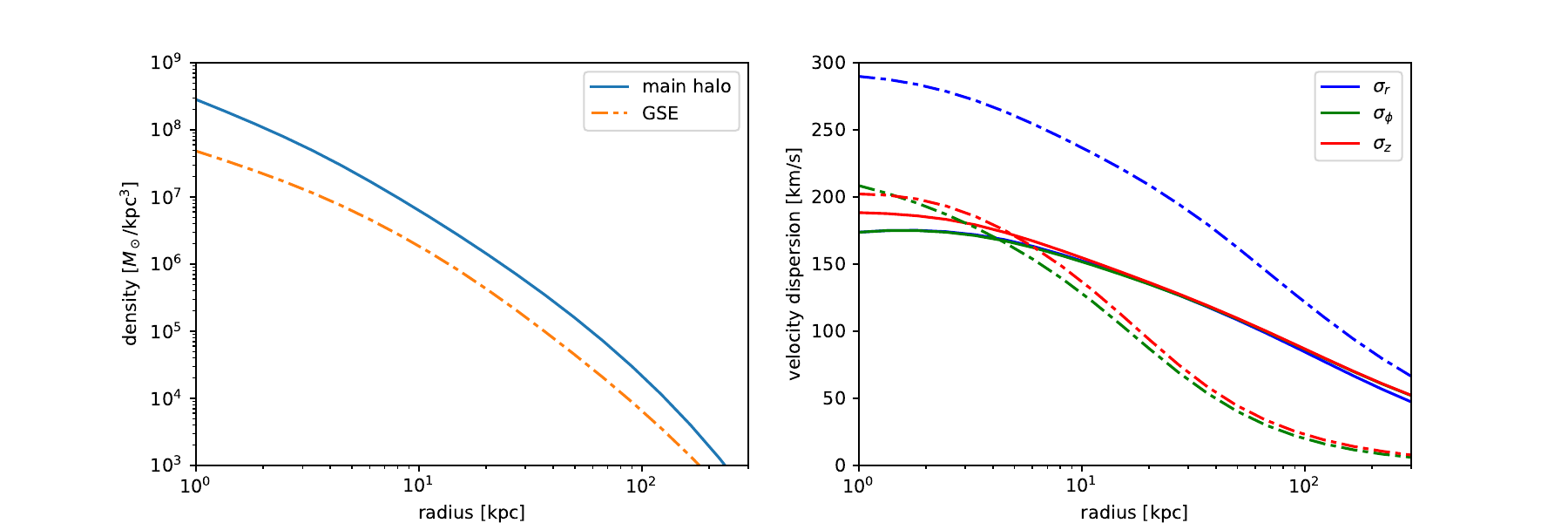}
    \caption{Mass density and velocity dispersion of the two scAHM components. The dispersion profile of the halo is close to isotropic in velocity space. The radial dispersion of the GSE component is much higher than the other two velocity components. }
    \label{fig:mass_vel_dispersion}
\end{figure*}

Despite numerous experiments, no significant signals or missing energies have been found. 
The only indications remain the previously mentioned astrophysical observations.

Taking these observations into account, one can infer the distribution of DM in the Milky Way galaxy and draw conclusions about its local density. 
This information can be used to estimate the magnitude of seasonal variations and directional effects, which could provide evidence for the discovery of DM \citep{earthmotion_spergel_1988}.
Seasonal variation is expected as Earth passes through varying densities of DM, depending on its relative velocity to the galactic center. 
In June, Earth's motion around the Sun aligns with the Solar System's motion, while in December, it opposes it. 
This would lead to a signal modulation with annual periodicity.

\section{Galactic halo models}

To estimate this modulation, a model of the local density and velocity distributions is needed.
A fundamental framework is provided by the standard halo model (SHM) \citep{halomodel_drukier_1986}, which posits that the Milky Way is embedded in a roughly spherical and isotropic DM halo extending beyond the visible components of the galaxy. 
\edit{However, the SHM has become outdated in multiple ways, particularly in its estimate of the local dark matter density, which was originally set at 0.3 GeV/cm$^3$ but has been revised upward to 0.4--0.5 GeV/cm$^3$ (see figure~1 in \cite{deSalas_2021} for a compilation of recent results).}
Recent observations from the Gaia space observatory have even revealed more complexity in our galaxy's halo structure, particularly with the discovery of the Gaia-Sausage-Enceladus (GSE) \citep{belokurov_2018,helmi_2018} -- an ancient merger between the Milky Way and a smaller dwarf galaxy, which likely occurred 8--10 Gyr ago and manifests itself as a population of halo stars with highly eccentric orbits.

In light of this discovery, \cite{evans_refinement_2019} introduced an updated DM model, referred to as SHM$^{++}$, which includes a DM counterpart to the GSE. 
Both the main component (nearly identical to SHM) and the GSE are assumed to follow smooth, spherically-symmetric density profiles and have Gaussian velocity distributions truncated at the escape velocity $v_\mathrm{esc}$, but while the former is still isotropic, the latter is assumed to have the same high radial anisotropy as stars, i.e. the anisotropy coefficient $\beta \equiv 1 - \sigma_\mathrm{tan}^2 / (2\sigma_\mathrm{rad}^2) \simeq 0.9$. 
The circular speed $v_0$, the escape speed, the local DM density, the fractional contribution of GSE and its anisotropy were updated based on recent measurements, predominantly from Gaia.
They then derived the analytical velocity distribution and other observables from this model.

However, the truncated Gaussian velocity distribution in the SHM$^{++}$ model is not very realistic. Although the true distribution function (DF) $f({\boldsymbol x}, {\boldsymbol v})$ of the DM halo is not observationally constrained, the Jeans theorem stipulates that in a steady state, it must be a function of the integrals of motion $\mathcal I(\boldsymbol x,\boldsymbol v;\;\Phi)$ in the potential $\Phi$.
Recently, \cite{mwmodel_binneyvasiliev_2023} constructed a dynamically self-consistent model of the Milky Way constrained by Gaia data, which provides a plausible choice for such a DF.
In the present paper, we explore the implications of this model on the DF velocity distribution in the Solar neighbourhood.

\section{Methods}

The Milky Way model of \cite{mwmodel_binneyvasiliev_2023} relies upon the AGAMA software package for galactic dynamics \citep{vasiliev_agama_2019} and uses actions $\boldsymbol J$ as the integrals of motion and arguments of the DF. The model is primarily specified by the DFs of its stellar and DM components, and the gravitational potential $\Phi$ is determined iteratively and self-consistently to ensure that the model is in a steady state (see section~5 in \citealt{vasiliev_agama_2019} for a detailed description of the method). The focus of the \cite{mwmodel_binneyvasiliev_2023} model was on the kinematics of disc stars, and the DM halo was represented by a single, nearly isotropic component. For the present paper, we modify the halo DF so that it contains two components: the main, isotropic component has the same parameters as before, but its mass is reduced by 20\%, and the remaining contribution comes from the GSE. Both components have the \texttt{DoublePowerLaw} DF \citep{posti_2015}:
\begin{eqnarray}  \label{eq:dm_df}
f(\boldsymbol J) &=& A\;
\left[1 + \frac{J_0}{h(\boldsymbol J)}\right]^\Gamma\;
\left[1 + \frac{g(\boldsymbol J)}{J_0}\right]^{\rm B}\;
\exp\left[-\left(\frac{g(\boldsymbol J)}{J_{\rm cut}}\right)^2\right] \nonumber\\
h(\boldsymbol J) &\equiv& k_{r,\rm in }\;\,J_r + k_{z,\rm in }\;\,J_z + (3-k_{r,\rm in }\;\,-k_{z,\rm in })\;\,|J_\phi|, \\
g(\boldsymbol J) &\equiv& k_{r,\rm out}J_r + k_{z,\rm out}J_z + (3-k_{r,\rm out}-k_{z,\rm out})|J_\phi|, \nonumber
\end{eqnarray}
and their parameters are given in table~\ref{tab:components} (the normalization factor $A$ is computed from the prescribed total mass). The remaining parameters describing the stellar components are the same as in the original paper. In what follows, we refer to this updated Milky Way model with a two-component DM halo as the self-consistent Anisotropic Halo Model (scAHM). \edit{We stress that while the DF of the GSE component needs to have strong radial anisotropy, there is no fundamental reason why it should follow the above functional form; here we chose it purely for convenience, as one of the standard models implemented in AGAMA. Ideally, one should look to cosmological simulations of Milky Way analogues with GSE-like merger history and find an appropriate form for the DF of the accreted component. \citet{Lane_2025} analysed an ensemble of Milky Way analogues from the IllustrisTNG simulation and concluded that DFs with a constant or radially increasing anisotropy of the Osipkov--Merritt form \citep{Cuddeford_1991} can describe the merger debris well; however, they only examined the velocity dispersion profiles, not the specific functional form of these DFs.}

Figure~\ref{fig:mass_vel_dispersion} shows the radial profiles of density and three components of velocity dispersion along the $x$ axis for both DM halo components: main halo by solid lines and GSE by dashed lines. The density profiles are designed to be similar at all radii, with the GSE component being 4$\times$ less massive. The velocity dispersion of the main halo is nearly isotropic, while GSE has $\sigma_r$ much higher than both tangential velocity components. \edit{However, the level of anisotropy in our model ($\beta \simeq 0.65$) is not as high as inferred from observations, which may be a limitation of our chosen DF family. Thus our conclusions about the effect of radial anisotropy on DM detection can only be regarded as qualitative.}

\begin{table}
\centering
\caption{Parameters for the DM halo components}
\begin{tabular}{lcc}  
\hline
parameter          & main & GSE \\
\hline
type               & \multicolumn{2}{c}{DoublePowerLaw} \\
mass / $10^{12}\,M_\odot$ & 0.8 & 0.2 \\
$\log_{10}(J_0$\quad    / [kpc\,km\,s$^{-1}$]) & 3.4 & 4.8 \\
$\log_{10}(J_{\rm cut}$ / [kpc\,km\,s$^{-1}$]) & 4.4 & 3.8 \\
$\Gamma$           & 1.3 & 1.1 \\
${\rm B}$          & 2.3 & 2.3 \\
$k_{\text{r,in}}$  & 1.4 & 0.1 \\
$k_{\text{r,out}}$ & 1.2 & 0.2 \\
$k_{\text{z,in}}$  & 0.8 & 1.4 \\
$k_{\text{z,out}}$ & 0.9 & 1.4 \\
\hline
\end{tabular}
\label{tab:components}
\end{table}

Using these DFs, we compute one-dimensional Galactocentric velocity distribution functions $f_i(v_i),\; i\in\{r,\phi,z\}$ for the radial, azimuthal and vertical velocity components at the solar location. 
To convert these into the velocity distribution $f_i(\boldsymbol u_i)$ in the heliocentric frame, one needs to subtract the Solar velocity $\boldsymbol v_\odot = \{-13,-246,8\}$~km\,s$^{1}$ \citep{drimmel_poggio_2018}. In addition to 1d projections, the full 3d velocity distribution $f(\boldsymbol u)$ can be visualized in spherical coordinates: the direction is plotted on the celestial sphere using the Hammer projection in different slices of the velocity magnitude $|\boldsymbol u|$.
Finally, for estimating the event rates in laboratory experiments, one needs the velocity $\boldsymbol w \equiv \boldsymbol v - \boldsymbol v_E(t)$ in the Earth (geocentric) reference frame, where $\boldsymbol v_{\text{E}}$ is the Earth velocity in the Galactocentric frame, which exhibits seasonal variations as it moves around the Sun.

Most detectors are sensitive only to the velocity magnitude (i.e. speed) $|\boldsymbol w|$, not to its direction. A general formula for the differential scattering rate $R$ of nuclear recoil events as a function of nuclear recoil energy $E_r$ \citep{recoil_smith_2001} is
\begin{equation}\label{eq:dRdE}
    \frac{dR(t)}{dE_r} = N_T \frac{\rho_0}{m_\chi} \int_{|\boldsymbol w|>w_{\text{min}}}\hspace{-5mm}
    |\boldsymbol w|\,f\big( \boldsymbol w + \boldsymbol v_{\text{E}}(t) \big)\, \frac{d\sigma_T(w,E_r)}{dE_r} d^3\boldsymbol w.
\end{equation}
In this context, $N_T$ denotes the number of target nuclei, $\rho_0$ represents the local DM density, and $m_\chi$ is the assumed DM particle mass. The minimum DM speed $w_{\text{min}}$ required to induce a recoil with energy $E_r$, is given by
\begin{equation}\label{eq:v_min}
    w_{\text{min}} = \sqrt{\frac{m_N E_r}{2\mu_N^2}}
\end{equation}
with nucleus mass $m_N$ and its reduced mass with the DM particle $\mu_N$.
The factor $\sigma_T$ refers to the DM nucleus scattering cross-section 
\begin{equation}\label{eq:crosssection}
    \frac{d\sigma_T(v, E_r)}{dE_r}=\frac{m_N A^2 \sigma_p^{\text{SI}}}{2\mu_p^2 v^2} F^2(E_r),
\end{equation}
where $A$ is the atomic number, $\sigma_p^{\text{SI}}=10^{-46}$ cm$^2$ is the assumed DM particle-proton cross-section, and $\mu_p$ is the proton-DM particle reduced mass. We employ the Helm nuclear form factor \citep{Feldstein_2010} for $F^2(E_r)$,
\begin{eqnarray}
    F^2(E_r)=\left( \frac{3j_1(qr_0)}{qr_0} \right)^2e^{-s^2q^2}.
\end{eqnarray}
$q=\sqrt{2 m_N E_r}$ is the momentum transfer, $r=1.2A^{1/3}$ fm is the nuclear radius, $r_0=\sqrt{r^2-5s^2}$ is the effective nuclear radius, $s=1$ fm is the nuclear skin thickness, and $j_1$ is the first spherical Bessel function.
%Lastly, $f(\vec{v}+\vec{v_E}(t))$ describes the DM velocity distribution retrieved in the previous step.
We calculate the expected event rates by integrating the scattering rate over the recoil energies.\\
Table \ref{tab:basic_properties} summarises basic properties of the SHM and scAHM model.

\begin{table}[h]
\centering
\caption{Basic properties of SHM and scAHM}
\renewcommand{\arraystretch}{1.7} % Adjust row spacing
\begin{tabular}{lcr}  
\hline
\multicolumn{3}{c}{SHM} \\
\hline
local density     & $\rho_0^\text{SHM}$        & 0.3 GeV/cm$^3$\\
circular velocity & $v_0$           & 220.0 km/s \\
escape velocity   & $v_\text{esc}$  & 542.5 km/s \\
% \hline\\
\hline
\multicolumn{3}{c}{scAHM} \\
\hline
local density             & $\rho_0^\text{scAHM}$                    & 0.45 GeV/cm$^3$\\
escape velocity           & $v_\text{esc}$                           & 542.5 km/s \\
halo density              & $\rho_0^\text{halo}$                     & 0.35  GeV/cm$^3$\\
GSE density               & $\rho_0^\text{GSE}$                      & 0.10  GeV/cm$^3$\\
halo anisotropy parameter & $\beta^\text{halo}$                      & -0.013\\
GSE anisotropy parameter  & $\beta^\text{GSE}$                       & 0.655\\
\hline
\end{tabular}
\label{tab:basic_properties}
\end{table}

\section{Results and discussion}
\begin{figure}
    \centering
    \includegraphics[width=1\linewidth]{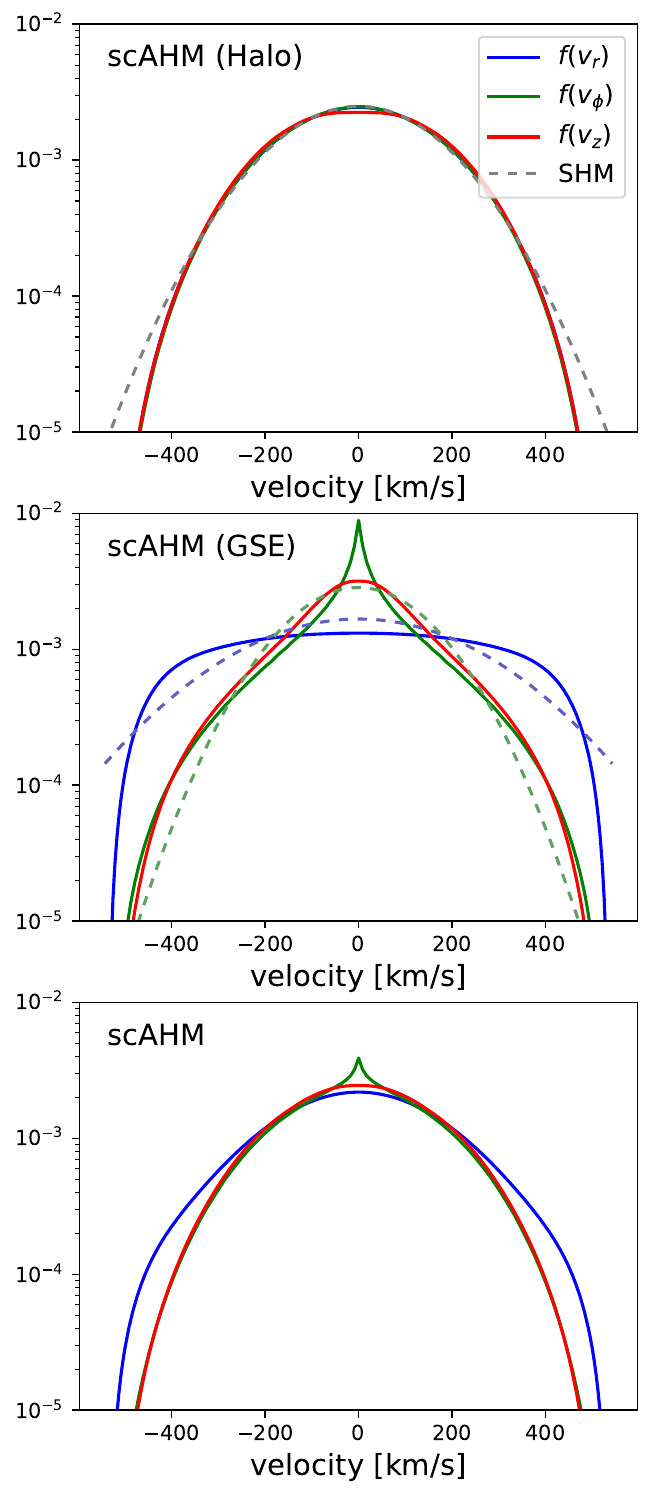}
    \caption{Velocity distribution functions $f(\vec{v})$ in the galactocentric reference frame. From top to bottom, we show the halo and GSE components of the scAHM model and the full scAHM model (the sum of the first two panels, normalized by their respective contributions of 80\% and 20\%). The top panel includes the Gaussian distribution of the SHM model (dashed line), while the middle panel shows the Gaussian profiles with the same velocity dispersions as $f(v_r)$ and $f(v_\phi)$ of the GSE component, illustrating their strong deviations from the normal distribution. The different colours represent the different directional components.\\}
    \label{fig:VDF} 
\end{figure}

\begin{figure*}
    \centering
    \includegraphics[width=1\linewidth]{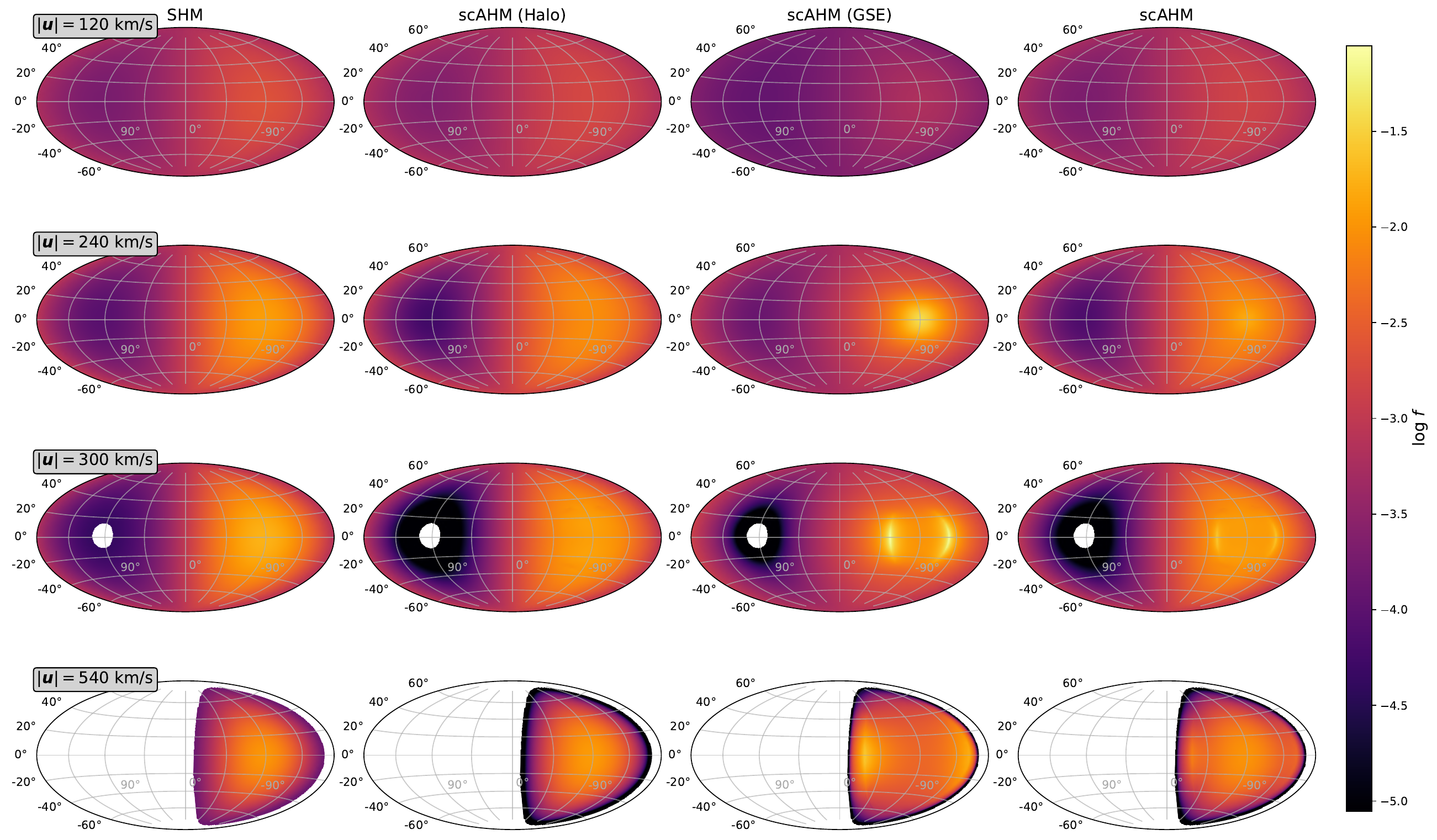}
    \caption{Sky maps of DM distribution functions in the heliocentric reference frame. The colour intensity represents the value of $f(\boldsymbol u)$ for particles with the given velocity magnitude (i.e. speed) $|\boldsymbol u|$, which increases from top to bottom, and the directional dependence is shown on the sky plane, where the direction to the Galactic centre is at origin, and the direction of Solar motion is in the centre of the left half of each panel. Qualitatively it can be interpreted as the expected signal in direction-sensitive laboratory experiments.
    The four columns represent different halo models, starting from left to right: SHM, Halo component of the scAHM model, GSE component of the scAHM model, and the full scAHM model. }
    \label{fig:skymaps}
\end{figure*}

\begin{figure*}
    \centering
    \includegraphics[width=1\linewidth]{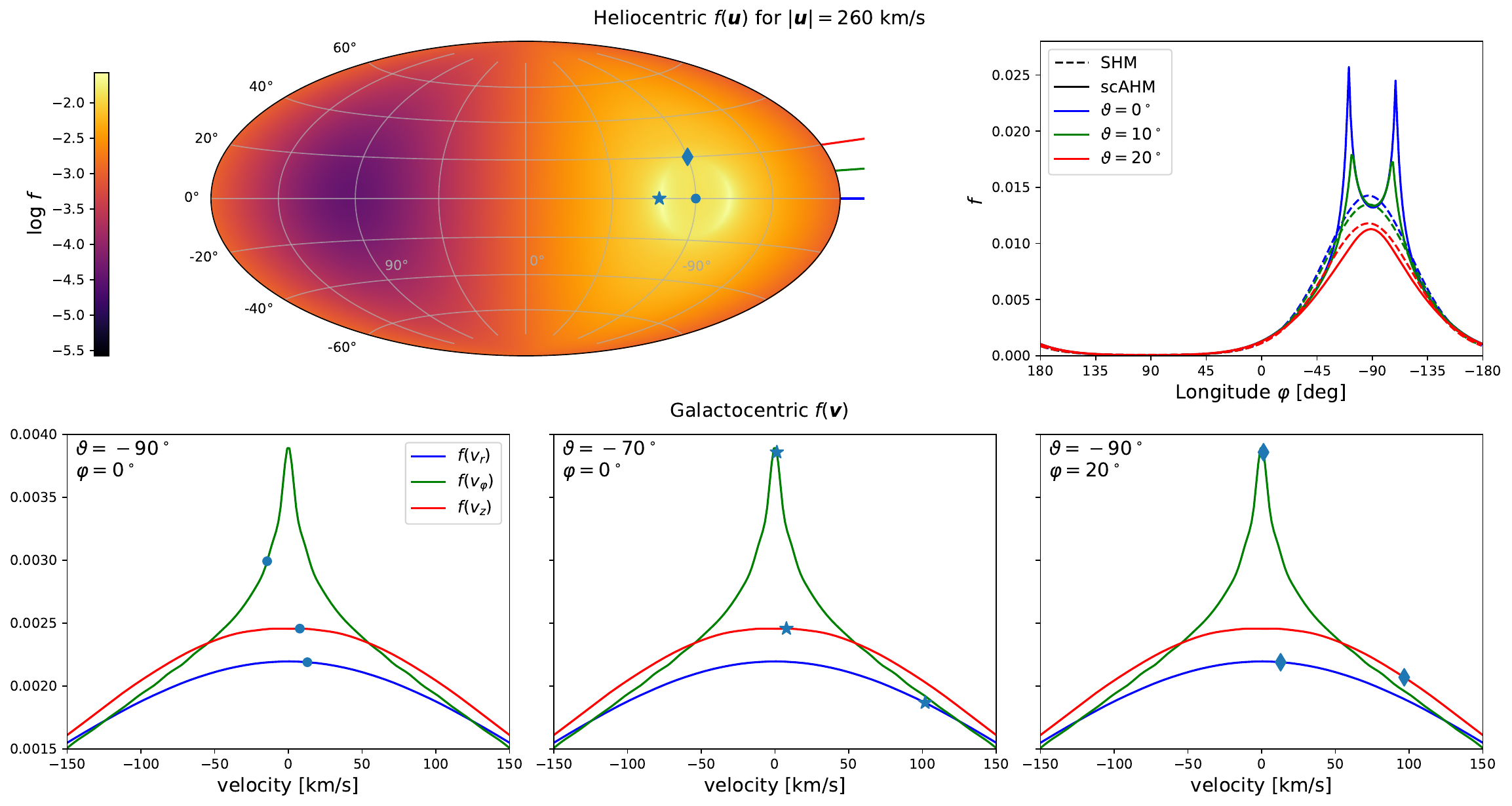}
    \caption{\textbf{Top left:} Sky map of $f(\boldsymbol u)$ for the scAHM model for the value of heliocentric speed $|\boldsymbol u|=260$~km\,s$^{-1}$ (roughly between the second and third rows of the right column of Figure~\ref{fig:skymaps}).
    The three different marked points correspond to particles moving in this direction with the given velocity as seen from the Sun. Note the Sun moves towards 90$^\circ$ longitude and 0$^\circ$ latitude.\\
    \textbf{Top right:} Horizontal slices of the same map at three values of latitude $\vartheta=\{0^\circ,10^\circ,20^\circ\}$, indicated by the same colours in the left panel. The scAHM DF is shown by solid lines, and the SHM DF -- by dashed lines.\\
    \textbf{Bottom row:} 1d Galactocentric velocity distributions of the scAHM model (same as in bottom panel of Figure~\ref{fig:VDF}). We marked the distribution function values on the three panels corresponding to the marked points in the skymap, although note that the full 3d DF is not identical to a direct product of its 1d projections.
    }
    \label{fig:VDF_and_Skymap}
\end{figure*}

\subsection{Galactocentric velocity distributions}
Having established realistic parameters of the model we determined the velocity distribution function by comparing the radial, azimuthal and vertical directions for both components of scAHM: the main DM halo of the Milky Way itself, contributing 80\% of the local DM density, and the GSE contributing the remaining 20\%.

We show the velocity distribution function in the galactic reference frame for each spatial direction in Figure \ref{fig:VDF}. 

The velocity distribution of the main halo component of scAHM slightly differs from the Gaussian of the SHM, having a steeper decline and a smooth transition to zero at $v>v_{\rm esc}$ instead of a sharp truncation. 
On the other hand, the GSE component has a significantly non-Gaussian distribution in all three components. The radial velocity distribution is much wider than a Gaussian and has a distinctly top-hat shape, while still smoothly transitioning to zero at the escape velocity. By contrast, the distributions of azimuthal and to a lesser extent vertical velocities have a narrower central peak than a Gaussian, and therefore a much smaller dispersion than the radial component. This is a natural consequence of the imposed radial anisotropy of this component: most of the kinetic energy is contained in the radial motion, thus the other two components are narrowly peaked around zero. The shape the velocity distribution is the most important difference between our model and SHM$^{++}$, which also includes a radially-anisotropic GSE component, but retains a truncated Gaussian approximation to its velocity distribution.
\edit{We note that the narrow spike in $f(v_\phi)$ (green curve) at zero might be an unphysical feature of our chosen DF family; however, in the case of a \citet{Cuddeford_1991} DF used by \citet{Lane_2025}, the spike is even more pronounced and manifests itself in both $v_\phi$ and $v_z$ components, although this hardly makes this DF any more realistic than our choice. Whatever the actual DF of the GSE component is, it likely has a peak (but not necessarily a spike) at small $v_\phi$ and $v_z$, perhaps lying somewhere between the red and green curves in our figure. }

\subsection{Heliocentric skymaps and their interpretation}

Figure~\ref{fig:skymaps} shows the full velocity distributions $f(\boldsymbol u)$ in the heliocentric reference frame in several slices of velocity magnitude $|\boldsymbol u|$ with varying velocity direction in the heliocentric reference frame shown as sky maps. 
The four columns represent different models, including SHM, the scAHM halo component, the scAHM GSE component and the full scAHM model.

In all panels, the dominant feature is the dipole anisotropy, which is a expected characteristic due to the the relative motion of our solar system. In this projection, the Sun moves in the direction of latitude $\varphi\approx 90^\circ$, i.e. the centre of the left half of each map. Thus particles arriving from that direction with a heliocentric speed $u$ have a Galactocentric speed $u+v_\odot$, whereas particles from the opposite direction $\varphi=-90^\circ$ have a lower Galactocentric speed $|u-v_\odot|$, at which the DF value is higher.

In the last two columns involving the GSE component, we additionally find an enhancement of signal around the direction opposite to the Solar velocity ($\varphi=-90^\circ$).

This feature first appears as a hot spot when the speed $u$ becomes comparable with the Solar velocity ($|\boldsymbol v_\odot| \simeq 250$~km\,s$^{-1}$), and as the speed increases, it forms a widening ring-like structure.

To better illustrate it, in the top row of Figure~\ref{fig:VDF_and_Skymap} we show the sky map of the scAHM model at $|\boldsymbol u|=260$~km\,s$^{-1}$, slightly above the Solar velocity, and several isolatitude slices $\vartheta=0^\circ,10^\circ,20^\circ$ of the same map. We mark three points on this sky map, and in the bottom row, show their location on the 1d Galactocentric velocity distributions $f(v_i)$, same as in the bottom row of Figure~\ref{fig:VDF}.
Location~1 (dot; bottom left panel) is in the centre of the ring shape and opposite of the direction of Solar motion. Therefore the respective velocity in the Galactocentric reference frame $\boldsymbol v$ has components $\{-13,14,8\}$~km\,s$^{-1}$. Because the velocity distribution in $v_\phi$ (green) is sharply peaked, even a small offset in $v_\phi$ from zero significantly decreases the DF value.
The other two locations are placed such that the azimuthal component $u_\phi$ exactly cancels the Solar azimuthal velocity $v_{\odot,\phi}$, so that $v_\phi\approx 0$ and the corresponding DF value is much higher. The difference between location~2 (star; bottom centre panel) and 3 (diamond; bottom right panel) is that $f(v_r)$ is much flatter for the GSE component than $f(v_z)$, therefore the DF value is higher for the same relative offset of $\sim 100$~km\,s$^{-1}$.

The geocentric velocity distributions are qualitatively similar to the ones shown in the above figures, but exhibit seasonal variations with an amplitude $\sim 30$~km\,s$^{-1}$.

\subsection{Expected Event Rates}
\begin{figure*}
    \centering
    \includegraphics[width=0.8\linewidth]{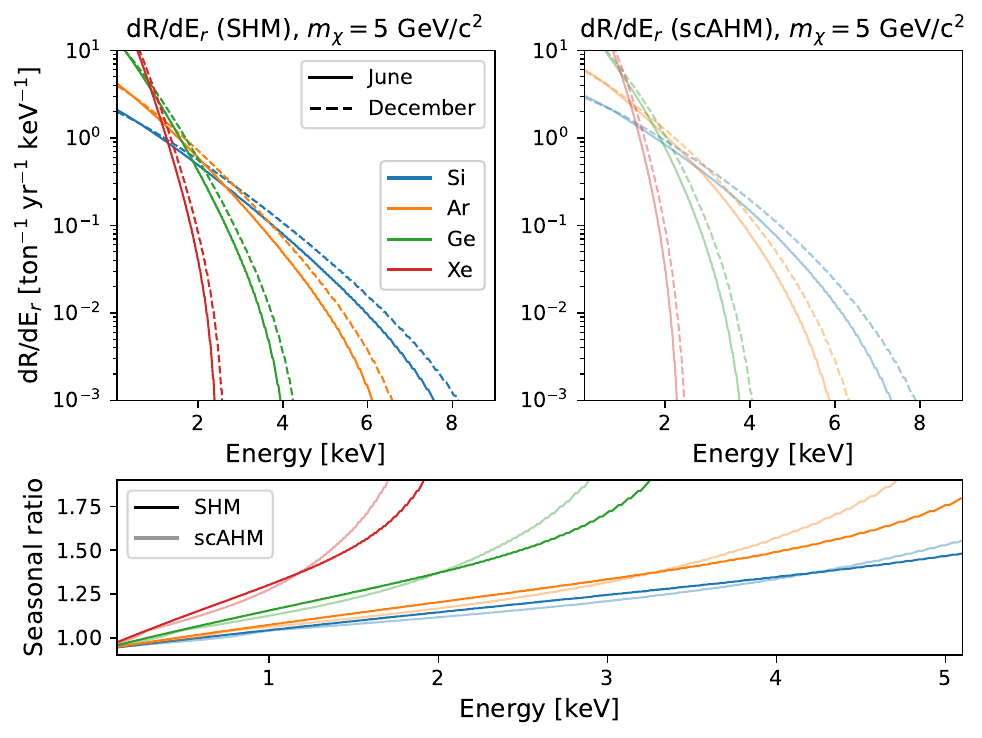}
    \caption{Differential event rates (top panels) and seasonal modulation (bottom panel) as a function of energy for WIMP experiments. The top left panel shows the SHM model while the top right panel shows the scAHM model. Different target nuclei (Si, Ar, Ge, Xe) are shown in different colours. The WIMP mass is assumed to be $m_\chi = 5$ GeV/c$^2$. Solid and dashed lines correspond to the seasonal variation for the expected rates in June and December, respectively. The bottom panel shows the seasonal ratio (solid versus dashed) of the two top panels. }\vspace{0.4cm}
    \label{fig:diffcrosssec}
\end{figure*}

We determined the differential event rates for the found velocity distribution functions using the previously introduced equation~\ref{eq:crosssection}. 
In Figure~\ref{fig:diffcrosssec} we compare the event rates and seasonal modulation for a spin-independent weakly interacting massive particle (WIMP) candidate in direction detection experiments under the two halo models.

As target nuclei, we chose Silicon, Argon, Germanium, and Xenon in a hypothetical experiment. 
For simplicity, we assumed the dark matter candidate to have a mass of $m_\chi = 5$ GeV/c$^2$.
The plots for the WIMP candidates with higher masses can be found in the appendix.

The event rate decreases with increasing recoil energy for all target nuclei and WIMP masses, reflecting the typical energy spectrum of WIMP interactions. 
The inclusion of the GSE component results in an elevated event rate especially at low recoil energies in comparison to the standard halo model in isolation.
This is expected as the GSE's narrower velocity distribution leads to a suppression of high-energy recoils. 
Heavier nuclei exhibit higher differential event rates than lighter nuclei for the same WIMP mass, due to their larger cross-section. 
A comparison of the seasonal dependence of the two models shows no significant change.

To quantify the differences between models, seasons and higher WIMP masses, we integrated the recoil energies (from 0 kEV to 100 keV) to determine the events expected in a year. For simplicity, we assume there is no lower energy threshold and all events are detected.
In table \ref{tab:events} we show the event rates for Xenon, which has the highest number of events and seasonal variation of all target nuclei.

\begin{table}
    \caption{Integrated event rates for Xenon}
    \label{tab:events}
    \centering
    \begin{tabular}{c c c c c c c c}
    \hline
    {m$_\chi$ [GeV/c$^2$]} & \multicolumn{2}{c}{$R_{\text{SHM}}$ [1/yr/ton]} & \hspace{5mm} & \multicolumn{2}{c}{$R_{\text{scAHM}}$ [1/yr/ton]} \\
    & June & December & & June & December \\
    \hline
    5   & 14.5 & 13.3  & & 23.3 & 21.5 \\
    20  & 37.7 & 36.12 & & 58.8 & 56.6 \\ 
    100 & 24.4 & 25.1  & & 36.3 & 37.3 \\
    \hline
    \end{tabular}
\end{table}

For all WIMP masses, the integrated event rates are slightly higher for the scAHM model than for SHM. 
This is consistent with the previous observation: the inclusion of the GSE leads to a modest increase in the overall scattering rate.

Neither season has a higher recoil rate for all energies. The earth moving with or against the stream of dark matter particles results in a surplus or deficit respectively of lower energy recoil events, but a deficit and a respective surplus of higher energy recoil events. In addition, the precise form of the differential recoil rate function varies significantly for different dark matter masses due to the Helm form factor and cutoff minimum velocity in the integration. Both of these effects combine to yield the behavior visible in Table \ref{tab:events} for the total integrated recoil rate for different dark matter masses and different seasons.

\section{Conclusion}
This study investigated the impact of the GSE merger event on the local DM velocity distribution and its implications for direct detection experiments. 
Utilizing the AGAMA framework and incorporating parameters derived from Gaia observations, we constructed self-consistent galaxy models including the GSE.

We demonstrate that the GSE introduces notable deviations from the smooth, Gaussian velocity distribution predicted by the SHM. 
The presence of radially anisotropic debris from the merger results in a narrower velocity distribution in the azimuthal and vertical direction and a broader distribution in the radial direction. 
These features manifest as distinct patterns in the simulated all-sky maps of recoil directions. Notably, the high-amplitude spike and ring in the direction opposite to the Galactocentric velocity of the Earth is a direct consequence of the cuspy velocity distribution in the azimuthal direction; such a signature is not found in the SHM$^{++}$ model, which approximates the velocity distribution of the GSE as an anisotropic Gaussian. However, revealing this feature requires a direction-sensitive detector.

Studying the effect of the GSE component on the differential event rates, we observe a slight decrease in differential event rates at high recoil energies, which is counterbalanced by a modest increase in overall differential event rates due to changes in other parameters. 

Furthermore, we see an overall increase of about 40$\%$ more events to be detected in the scAHM model compared to the SHM model.
\edit{This is mainly driven by a comparable increase of the local density from 0.3 to 0.45 GeV/cm$^3$ (Table~\ref{tab:basic_properties}), and less by the anisotropic velocity distributions.}
The exact number of detected events depends strongly on the model assumptions and was calculated assuming an efficiency of unity; therefore, the quantitative values should be interpreted with caution. \edit{Moreover, the details of the velocity distribution (Figure~\ref{fig:VDF}) and the directional dependence of the signal (Figure~\ref{fig:VDF_and_Skymap}) are determined by the specific functional form of the DF of the accreted component; here we examined one possible choice, but a more comprehensive analysis of DFs found in cosmological simulations is beyond the scope of this study. It is unlikely that the overall conclusions about detection rates would be strongly affected by our DF choice, although it may be important for direction-sensitive laboratory experiments. }

A limitation of our model is the exclusion of potential DM halo perturbations from external galaxies, notably the Large Magellanic Cloud (LMC). 
Simulations \citep[e.g.,][]{Besla_2019, Donaldson_2022, SmithOrlik_2023} suggest that the LMC could induce a high-speed tail in the local DM velocity distribution. 
While a full consideration is beyond our scope, we justify this exclusion by assuming that the GSE merger dominates the non-Gaussian features, with the LMC's influence representing a second-order correction.

Altogether, the impact of the GSE component would be most clearly observable in direction-sensitive detectors, where its distinct velocity distribution features could be directly measured.

\section*{Acknowledgments}
We would like to thank Laura Baudis and the anonymous referee for their insightful comments and valuable suggestions, which greatly helped to improve our paper. EV acknowledges support from an STFC Ernest Rutherford fellowship (ST/X004066/1).

\bibliographystyle{aa}
\bibliography{references}

\appendix

\begin{figure*}
    \centering
    \includegraphics[width=0.8\linewidth]{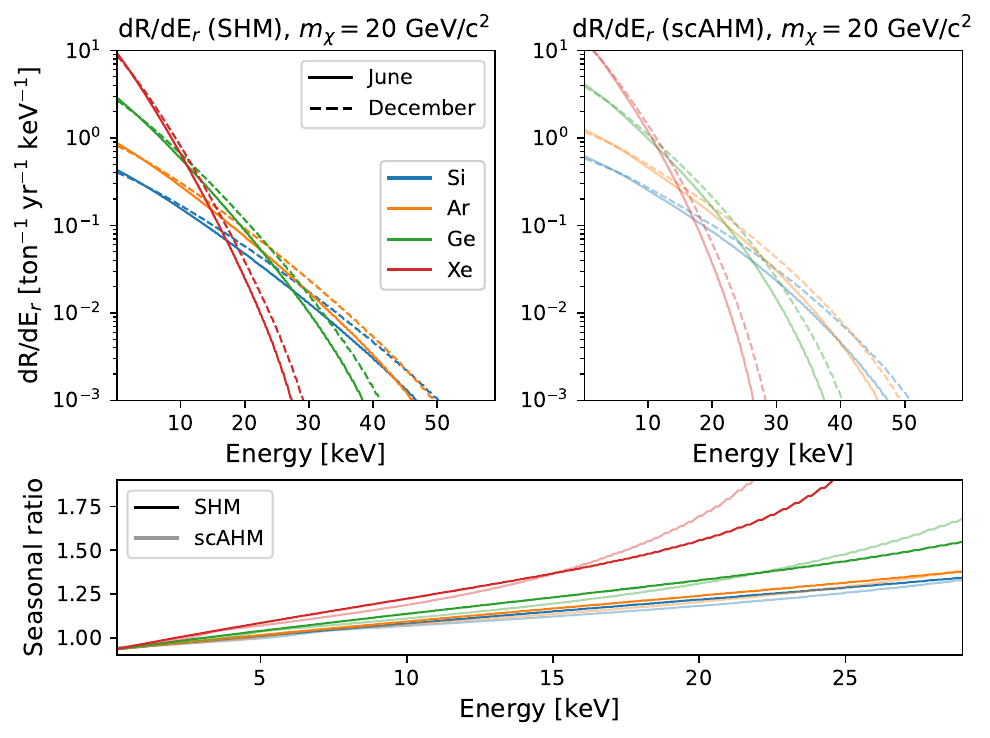}
    \caption{Like Figure \ref{fig:diffcrosssec}, but for a WIMP mass of $m_\chi = 20$ GeV/c$^2$.}\vspace{0.4cm}
    \label{fig:diffcrosssec1}
\end{figure*}

\begin{figure*}
    \centering
    \includegraphics[width=0.8\linewidth]{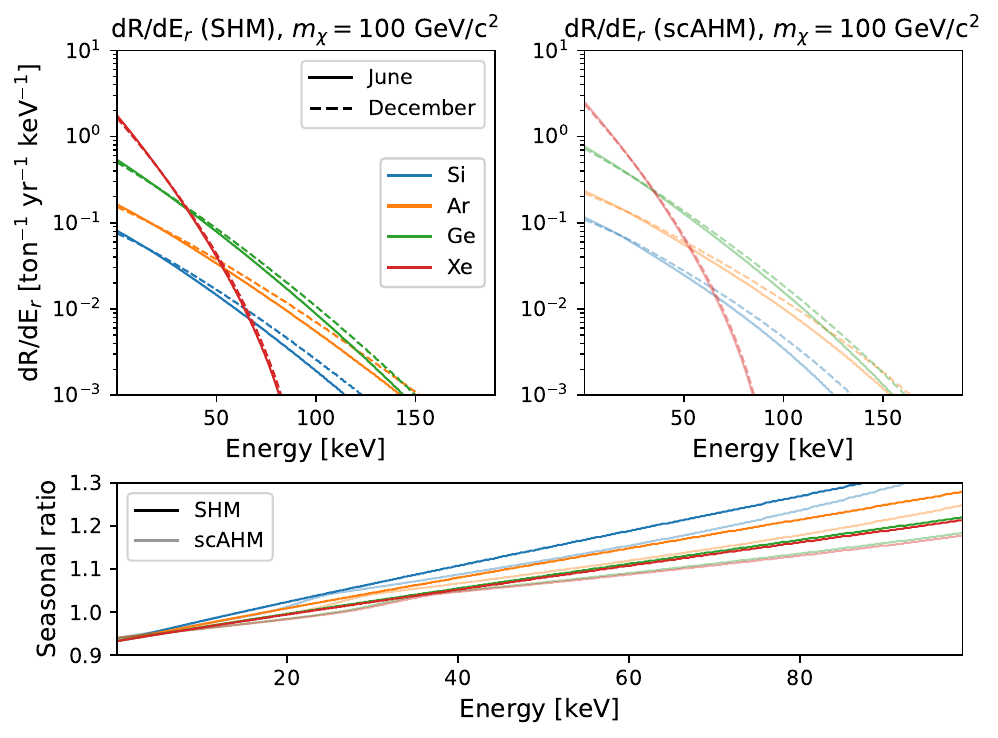}
    \caption{Like Figure \ref{fig:diffcrosssec}, but for a WIMP mass of $m_\chi = 100$ GeV/c$^2$.}\vspace{0.4cm}
    \label{fig:diffcrosssec2}
\end{figure*}

We present higher WIMP mass analogues of Figure \ref{fig:diffcrosssec} in Figures \ref{fig:diffcrosssec1} and \ref{fig:diffcrosssec2}.

\newpage

\end{document}